\documentclass[onecolumn,11pt,preprintnumbers,amsmath,amssymb]{revtex4}

\usepackage{epsf}
\usepackage{graphicx}  
\usepackage{dcolumn}   
\usepackage{bm}        

\newcommand{\be}{\begin{equation}}
\newcommand{\en}{\end{equation}}
\newcommand{\ba}{\begin{array}}
\newcommand{\ea}{\end{array}}
\newcommand{\bea}{\begin{eqnarray}}

\newcommand{\ena}{\end{eqnarray}}

\begin{document}

\preprint{arxiv:}

\title{Brane tachyon dynamics}

\author{Hongsheng Zhang \footnote{Electronic address: hongsheng@kasi.re.kr}}
\affiliation{Shanghai United Center for Astrophysics (SUCA),
 Shanghai Normal University, 100 Guilin Road, Shanghai 200234,
 P.R.China}
 \affiliation{
 Korea Astronomy and Space Science Institute,
  Daejeon 305-348, Korea }
 \affiliation{ Department of Astronomy, Beijing Normal University,
Beijing 100875, China}
   \author{Xin-Zhou Li \footnote{Electronic address: kychz@shnu.edu.cn} }
 \affiliation{Shanghai United Center for Astrophysics (SUCA), Shanghai Normal
University, 100 Guilin Road, Shanghai 200234, P.R.China}

 \author{Hyerim Noh\footnote{Electronic address: hr@kasi.re.kr} }
 \affiliation{
 Korea Astronomy and Space Science Institute,
  Daejeon 305-348, Korea }


\begin{abstract}
 The dynamics of a tachyon attached to a Dvali,
 Gabadadze and Porrati (DGP) brane is investigated. Exponential
 potential and inverse power law potential are explored,
 respectively. The quasi-attractor behavior, for which the universe will eventually go into a phase similar to the
 slow-roll inflation, is discovered in both  cases of exponential
 potential and  inverse power law potential. The equation
 of state (EOS) of the virtual dark energy for a single scalar can cross the phantom
 divide in the branch $\theta=-1$ for both  potentials, while the
 EOS of the virtual dark energy for a single scalar can not cross
 this divide in the branch  $\theta=1$.
\end{abstract}


 \maketitle

\section{Introduction}

$E_8\times E_8$ heterotic string emerges when one compacts the
11-dim
 super gravity on an $S^1/Z_2$ orbifold (Horava-Witten proposal) \cite{hw}. The behavior of the string theory in the energy region
  below the unification scale is not sensitive to the fine structure of the  inner
 Calabi-Yau space, that is, the universe can be effectively described by some 5-dim theory, in which
 the standard model particles are confined to the 3-brane, while the gravitation can propagate in
 the whole spacetime. Such string-inspired phenomenological model, called brane world, has been set up and studied extensively
   \cite{maarten}, especially in the fields of high energy physics and
  cosmology. According to their behavior in different energy scales, brane world models can be classified into two
  main categories. One is ``high energy theory", that is, its phenomenology becomes different in high energy (ultra-violet)
   region from general relativity (GR), but recovers to GR in low
   energy (infrared) region, for example Randall-Sundrum (RS) model \cite{rs}. On the contrary,
   the other type of brane world, ``low energy theory", concentrates on the modification in
   low energy region. Dvali,
 Gabadadze and Porrati (DGP) model \cite{dgpmodel} is a leading model in the
   low-energy-theory models, which is mainly applied to the late
   time universe (see, however, \cite{dgpinflation}). Great
   interest has been aroused in the researches of  late time universe since the discovery of cosmic acceleration.

  The present cosmic acceleration is one of the most significant cosmological discoveries
 over the last century \cite{acce}. The physical nature of this acceleration remains as a
 mystery. Various explanations have been proposed, such as a small positive
 cosmological constant, quintessence, k-essence, phantom, holographic dark energy,
 etc., see \cite{review} for
 recent reviews with fairly complete list of references of different models.
 A cosmological constant is a  simple candidate for dark
  energy. However, following the more accurate data a more dramatic result
  appears: the recent analysis of the type Ia supernovae data
  indicates that the time varying dark energy gives a better
  fit  than a cosmological constant, and in particular, the equation of state (EOS)
   parameter  $w$ (defined as the ratio of
 pressure to energy density) may cross the phantom divide $w=-1$ \cite{vari}.  Three roads to cross this divide were
 summarized
 in a recent review article
 \cite{reviewcross}: i). quintom type (two-field)  model, for a review see \cite{reviewquintom}, ii).
 interacting model, for example see \cite{self2}, and iii). model in frame of new gravity, especially
 brane world, for example, see \cite{self3}.

 Inspired by the hopeful unification theory, string/M theory, the models in frame of
 modified gravity are duly noted since they offer much more extensive possibilities for dark
 energy. A useful example is that a single scalar can not cross the
 phantom divide while it can cross the divide in frame of DGP \cite{self3}. Brane
 world model inherits a key geometric property of 11-dim the Horava-Witten proposal of string/M, which requires standard model
 particles confined to a brane, while gravity propagates freely throughout the whole manifold.
 On the other hand an exotic matter with negative pressure, tachyon, coming from string/M theory
 also has been widely applied in cosmology,  as inflaton in the early
 universe \cite{tachyonin}, and as dark energy in the late time universe \cite{tachyonde}. Thus, it is
 interesting to study the dynamics of a tachyon attached to a brane.

 Tachyon is a field at the top of its potential, which has a fairly
 long history in particle physics. It returns with the studies of
 string/M recently. It was found that the tachyon modes of open
 string attached to a Dp-brane described the inherent instability of
 the Dp-brane \cite{sen}. A tachyon field has negative pressure,
 therefore it may be a proper candidate to drive the universe to
 accelerate. Generally speaking, a tachyon is always associated to a
 brane. The behavior of a tachyon in RS type brane world has been
 investigated in \cite{tachyonRS}, which is concerned with the early universe. In this
 article, we will study the behavior of a tachyon field in DGP. We
 focus on the late time evolution of the universe in the present article. In the standard model (4-dim GR)
 , the equation of state (EOS) of a tachyon is always in the interval $(-1, 0)$. However, we will
 show that the effective EOS of the dark energy  in the tachyon-DGP model can
 cross the phantom divide, which satisfies the amazing possibility
 of the crossing behavior of dark energy implied by recent observations.

  To find an attractor solution is an important method in cosmology,
  which is helpful to alleviate the coincidence problem. If there
  does not exist an attractor in a system, the commonsensible lore
  tells
  us that the orbits of the phase portrait never converge: it will
  look like a turbulent flow. However, we discover the qusi-attractor in the
  dynamical system without any critical point. The orbits with
  different initial conditions will converge to a quasi-de Sitter
  evolution. A useful analogy of this quasi-attractor is the
  slow-roll inflation. We find that the quasi-attractor behavior is
  rather robust, which will appear in the cases of different
  potentials
  of a tachyon.

  The outline of this article is as follows. In the next section we
  present our set up of the model. In section III, we study the
  evolution of this system via a dynamical system analysis. In
  section IV, we conclude this article.

\section{The Model}
 Let's start from the action of the DGP model
 \be
 \label{totalaction}
 S=S_{\rm bulk}+S_{\rm brane},
 \en
where
 \be
 \label{bulkaction}
  S_{\rm bulk} =\int_{\cal M} d^5X \sqrt{-{}^{(5)}g}
  {1 \over 2 \kappa_5^2} {}^{(5)}R ,
 \en
and
 \be
 \label{braneaction}
 S_{\rm brane}=\int_{M} d^4 x\sqrt{-g} \left[
{1\over\kappa_5^2} K^\pm + L_{\rm brane}(g_{\alpha\beta},\psi)
\right].
 \en
Here $\kappa_5^2$ is the  5-dim gravitational constant, ${}^{(5)}R$
is the 5-dim curvature scalar. $x^\mu ~(\mu=0,1,2,3)$ are the
induced 4-dim coordinates on the brane, $K^\pm$ is the trace of
extrinsic curvature on either side of the brane and $L_{\rm
brane}(g_{\alpha\beta},\psi)$ is the effective 4-dim Lagrangian,
which is given by a generic functional of the brane metric
$g_{\alpha\beta}$ and matter fields $\psi$ on the brane.

 Consider the brane Lagrangian consisting of the following terms
\begin{eqnarray}
\label{lbrane}
 L_{\rm brane}=  {\mu^2 \over 2} R  + L_{\rm
m}+L_{T},
\end{eqnarray}
where $\mu$ is 4-dimensional reduced Planck mass, $R$ denotes the
curvature scalar on the brane, and $L_{T}$ represents the Lagrangian
of a tachyon attached to the brane, $L_{\rm m}$ stands for the
Lagrangian of other matters on the brane.
 Then, assuming a  mirror symmetry in the bulk, we
  have the Friedmann equation on the brane \cite{dgpcosmology},
    \bea
 H^2+\frac{k}{a^2}=\frac{1}{3\mu^2}\left[\rho+\rho_0+\theta\rho_0
 (1+\frac{2\rho}{\rho_0})^{1/2}\right],
 \label{fried}
 \ena
 where $H\triangleq \dot{a}/a$ is the Hubble parameter, $a$ is the
 scale factor, $k$ is the spatial curvature of the
  three dimensional maximally symmetric space in the FRW metric on the brane,
  and $\theta=\pm 1$ denote the two branches of DGP model, $\rho$ denotes
  the total energy density, including dust matter and tachyon, on the brane,
  \be
  \rho=\rho_{T}+\rho_{dm}.
  \en
  The term $\rho_0$ relates the the strength of the 5-dim
  gravity with respect to the 4-dim gravity,
 \be
 \rho_0=\frac{6\mu^2}{r_c^2},
 \en
 where the cross radius is defined as $r_c\triangleq \kappa_5^2\mu^2$.

  A no-go theorem shows that a single field with reasonable
  conditions in GR can not cross the phantom divide.
  We will show that in our model only one field  is enough for this crossing
  behaviour via the effect of the 5-dim gravity.
  In fact, the accelerated expansion of the universe is a joint effect of the tachyon and
  the competition between 4-dim gravity and the 5-dim gravity.

 In the brane world model,
   the surplus geometric terms relative to the Einstein tensor
   play the role of the dark energy in GR in part.
    However, almost all observed properties of dark energy are
    obtained in frame of GR with a dark energy.
    To explain the the observed evolving  EOS of the effective dark
    energy,  we introduce the concept ``equivalent
 dark energy" or ``virtual dark energy" in the modified gravity
 models \cite{reviewcross}.  We derive the density of virtual dark energy caused by the tachyon and induced gravity term
  by comparing the modified Friedmann equation in
  the brane world scenario with the standard Friedmann equation in general
  relativity.
   The Friedmann equation in the
 4-dimensional
  GR can be written as
 \be
 H^2+\frac{k}{a^2}=\frac{1}{3\mu^2} (\rho_{dm}+\rho_{de}),
 \label{genericF}
 \en
 where the first term of RHS in the above equation represents the dust matter and the second
 term stands for the dark energy. Comparing (\ref{genericF})
 with (\ref{fried}), one obtains the density of virtual dark
 energy of DGP,
 \be
 \rho_{de}=\rho_{T}+\rho_0+\theta\rho_0
 (1+\frac{2\rho}{\rho_0})^{1/2}.
 \label{rhode}
 \en

 Since the dust matter obeys the continuity equation
 and the Bianchi identity keeps valid, dark energy itself satisfies
  the continuity equation
 \be
 \frac{d\rho_{de}}{dt}+3H(\rho_{de}+p_{eff})=0,
 \label{contieff}
 \en
 where $p_{eff}$ denotes the effective pressure of the dark energy.
 And then we can express the equation of state for the dark
 energy as
   \be
  w_{de}=\frac{p_{eff}}{\rho_{de}}=-1-\frac{1}{3}\frac{d \ln \rho_{de}}{d \ln
  a}.
  \label{wde}
   \en
   Observing the above equation, we find that the behavior of $w_{de}$ is determined by the term $\frac{d \ln \rho_{de}}{d \ln
  a}$. $\frac{d \ln \rho_{de}}{d \ln
  a}=0$ (cosmological constant) bounds phantom and quintessence. More
  intuitively, if $\rho_{de}$ decreases and then increases, or increases and then
  decreases with the expansion of the universe, we are certain that EOS of dark energy crosses phantom
  divide. A more important reason why we use the density to describe
  property of dark energy is that the density is
  more closely related to observables, hence is more tightly
  constrained for the same number of redshift bins used \cite{wangyun}.

   \section{dynamics of tachyon-DGP}

   In this section, we will analyze the dynamics
   of a tachyon in the late time universe on
   a  DGP brane with two different potentials, respectively.
   We show that the quasi-attractor (which we will explain in detail later) appears in both of the two cases.

  For a tachyon field in a curved background, the action in $L_{T}$ of (\ref{lbrane})
  takes a Dirac-Born-Infeld (DBI) form,
  \be
  L_{T}=-V(T)\sqrt{1+X},
  \en
  where
  \be
  X=g^{\mu \nu}\partial _\mu T\partial _\nu T.
  \en
  One sees that a tachyon has a dimension of [length] rather than
  [mass], which is different from  an ordinary scalar.
  The equation of motion for tachyon reads,
  \bea
 \nonumber
 \frac{1}{V(T)}\frac{dV(T)}{dT}-(\frac{1}{\sqrt{-g}}\partial _{\mu}\sqrt{-g})
 g^{\mu\nu} \partial _{\nu}T+ \frac{1}{2(1+X)}g^{\mu\nu}
 \partial_{\nu}T (\partial_{\mu} g^{\alpha\beta}\partial_{\alpha}T
 \partial_{\beta}T+ 2g^{\alpha\beta}\partial_{\alpha} \partial_{\mu}T
 \partial_{\beta}T)
 \\
 -\partial_{\mu}g^{\mu\nu}\partial_{\nu}T- g^{\mu\nu}
 \partial_{\mu}\partial_{\nu}T=0,
  \label{eom}
 \ena
 which degenerates to
 \be
 \frac{\ddot{T}}{1-\dot{T} ^2}+3H\dot{T}+\frac{1}{V(T)}\frac{dV(T)}{dT}=0,
 \label{coseom}
 \en
 in an FRW universe, where a dot denotes the derivative with respect to time.
 Note that our result (\ref{eom}) is different from the result in
 \cite{gib}, which cannot degenerate to (\ref{coseom}) in an FRW
 universe.

  Varying the action with respect to the metric tensor  we
  obtain the energy momentum of the tachyon field,
  \be
  T^{\mu\nu}=-V\left[g^{\mu\nu}(1+X)^{1/2}-(1+X)^{-1/2}\partial^{\mu}T\partial^{\nu}T
  \right],
  \en
 which reduces to
   \bea
\rho &=& \frac{V(T)}{\sqrt{1-\dot{T}^2}}~, \\
p &=& -V(T)\sqrt{1-\dot{T}^2}~,
   \ena
   in an FRW universe. Thus the
  (local) equation of state of tachyon reads,
  \be
  w=\dot{T}^2-1.
  \en
  The reality conditions for $\rho$ and $\dot{T}$ require $0\leq
  \dot{T}^2\leq 1$, which yields,
  \be
  -1\leq w\leq 0.
  \en

  For a more detailed research of the evolution of the variables in this model
  we write them in a dynamical system, which can be derived from the Friedmann equation
  (\ref{fried}) and continuity equation (\ref{contieff}). We first define some new
  dimensionless variables,
  \bea
  x&\triangleq&\dot{T},\\
  y&\triangleq&\frac{\sqrt{V}}{\sqrt{3}\mu H},\\
  l&\triangleq&\frac{\sqrt{\rho_{dm}}}{\sqrt{3}\mu H},\\
  b&\triangleq&\frac{\sqrt{\rho_0}}{\sqrt{3}\mu H}.
  \ena
 The physical meanings of these new variables are clear: $x$ denotes
 kinetic energy of the tachyon, $y$ marks the relative strength of  potential energy
 to the Hubble parameter, $l$ represents the relative strength of
 the dust density  to the Hubble parameter, and $b$ stands for the Hubble parameter.
  The exact form of the potential of a tachyon is still under research. In the following two subsections, we will
  discuss two examples of potentials, say, exponential potential and
  inverse power law potential.

  \subsection{exponential potential}

    The exponential potential is an important
   example which can be solved exactly in the standard model for a
   scalar. We first study the dynamics of a tachyon with an
   exponential potential,

   \be
   V=V_0e^{-\lambda T},
   \en
  where $V_0$ and $\lambda$ are two constants. With the evolution of
  the universe, the tachyon rolls down, which can be described by,

  \bea
 \label{yi}
 x'&=&{3}(1-x^2)(-x+jb),\\
 \label{er}
  y'&=&\frac{3}{2}\alpha y-\frac{3}{2} jbxy,\\
 \label{sa}
   l'&=&\frac{3}{2}\alpha l-\frac{3}{2}l,\\
 \label{si}
  b'&=&\frac{3}{2}\alpha b,
 \ena
 where \be j=\frac{\mu \lambda}{\sqrt{3\rho_0}}, \en
 and
  \be
 \alpha\triangleq \left[l^2+x^2y^2(1-x^2)^{-1/2}\right]\left[1+\theta\left(1+2\frac{y^2(1-x^2)^{-1/2}+l^2}{b^2}\right)^{-1/2}\right],
 \label{alpha}
 \en
 and a prime stands for derivation with respect to
 $s\triangleq \ln a$. $\alpha$ has significant physical sense in a dynamical universe. In fact it is just the
 slow-roll parameter in the language of inflation,
 \be
 \alpha=-\frac{2\dot{H}}{3H^2}.
 \en
 $\alpha<<1$ implies the universe enters a quasi-de Sitter phase.

  In the above system we have set $k=0$, which is implied
 either by theoretical side (inflation in the early universe),
  or observational side (CMB fluctuations \cite{WMAP}). One can check this system degenerates to a
 tachyon with dust matter in standard GR.
  Note that the 4 equations (\ref{yi}), (\ref{er}),
 (\ref{sa}), (\ref{si}) of this system are not independent. By using the Friedmann
 constraint, which can be derived from the Friedmann equation,
 \be
 \label{constraint}
 y^2(1-x^2)^{-1/2}+l^2+b^2+\theta b^2\left(1+2\frac{y^2(1-x^2)^{-1/2}+l^2}{b^2}\right)^{1/2}=1,
 \en
 the number of the independent equations can be reduced to 3.
    There are two critical points of this system satisfying $x'=y'=l'=b'=0$ appearing at
 \bea
 &x&=y=l=b=0;\\
 &x&=1,~~y=l=b=0.
 \ena
 It is easy the check that neither of them satisfies the Friedmann constraint
 (\ref{constraint}). So there is no de Sitter type attractor in this system.

 However, through a detailed  numerical investigation, we
 find a future ``quasi-attractor" in this system.
 The physical picture is that for a fairly large space of the
 initial conditions, the tachyon on a DGP will enter such a quasi-de Sitter
 space, where the tachyon rolls down the potential very slowly such
 that its kinetic energy is effectively zero and the background space is in
  fact a de Sitter one. One can  make an analogy to
 the situation of slow-roll inflation, where we omit the kinetic
 energy of the inflaton and we treat the spacetime as a de Sitter.

\begin{figure}
\centering
 \includegraphics[totalheight=5.6in, angle=-90]{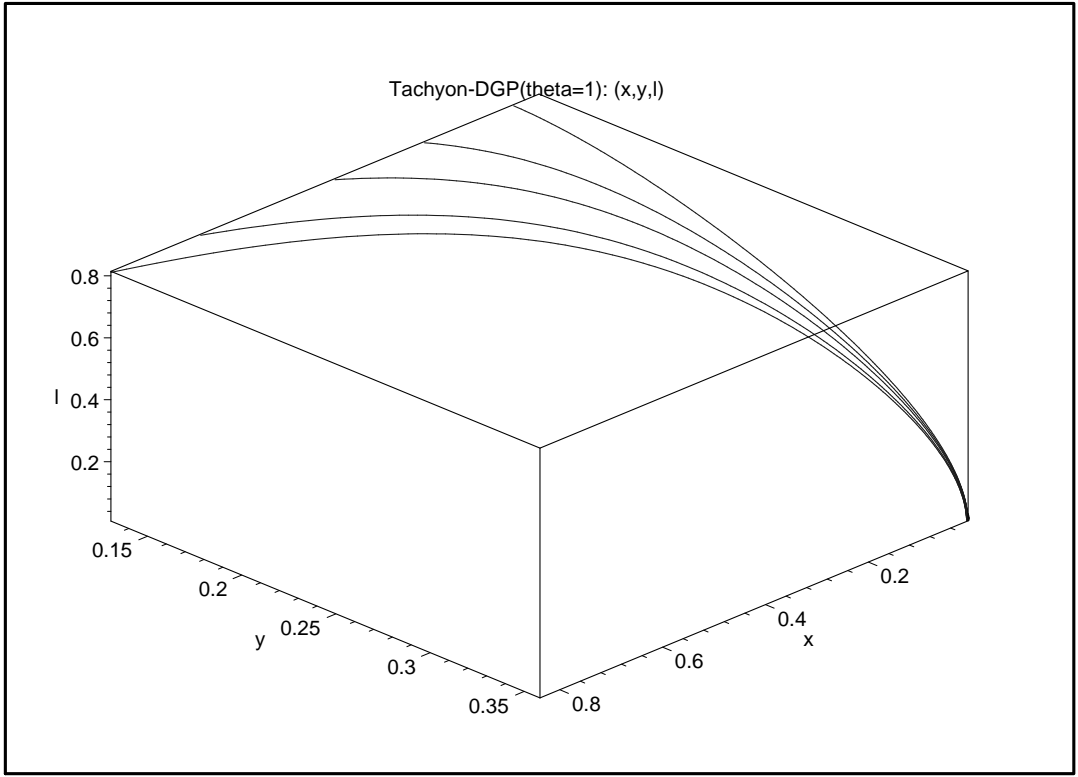}
\caption{The evolution of a tachyon on a DGP in the branch
$\theta=1$.  The different initial conditions for the curves are
$x(s=0)=0.1, x(s=0)=0.08, x(s=0)=0.05, x(s=0)=0.03, x(s=0)=0.01$
from the left to the right, respectively.}
 \label{l-x-y-exp}
 \end{figure}

\begin{figure}
\centering
 \includegraphics[totalheight=3in, angle=-90]{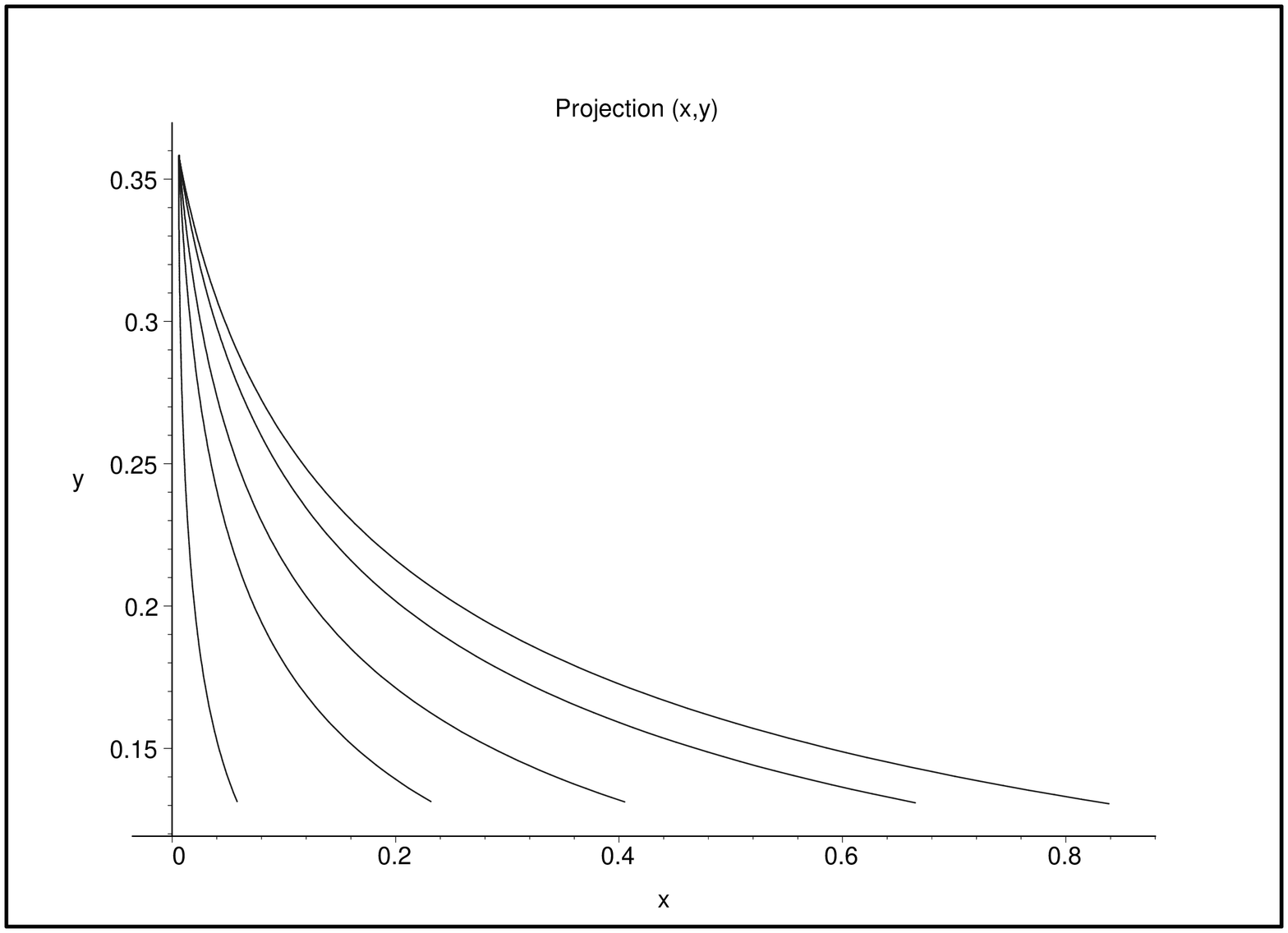}
 \includegraphics[totalheight=3in, angle=-90]{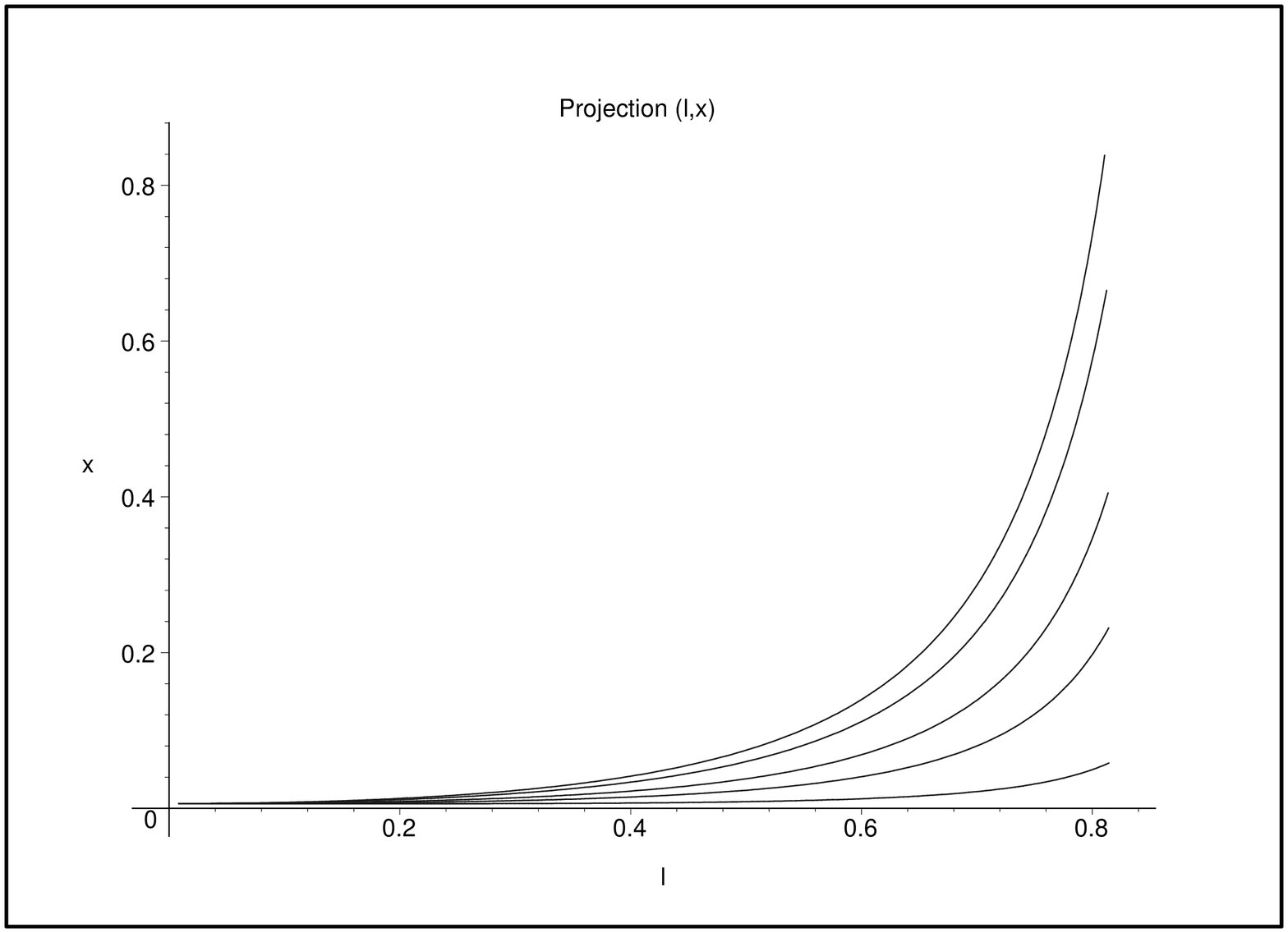}
 \includegraphics[totalheight=3in, angle=-90]{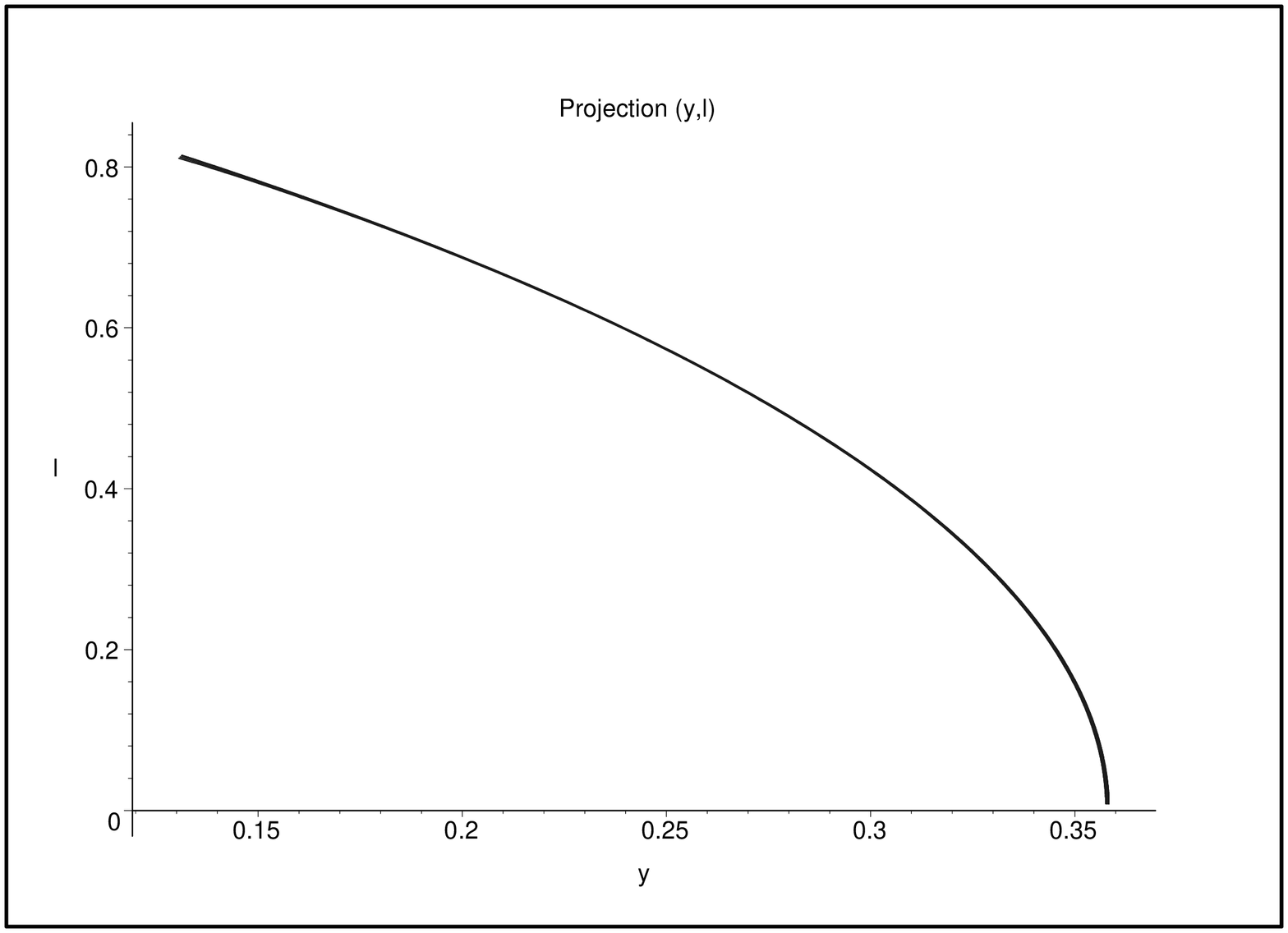}
\caption{The projections of fig \ref{l-x-y-exp} on $x-y$, $x-l$,
$l-y$ planes, respectively.}
 \label{projexp}
 \end{figure}

 It is difficult to obtain the analytical solution of the
 quasi-attractor. We show the typical obits of this system in the
 two branches, respectively. Since $l$ is an explicit function of
 $b$,
 \be
 l\sim ba^{-3/2}=be^{-3s/2},
 \en
 we just plot 3-dim phase portraits in the subspace $x-y-l$ in fig \ref{l-x-y-exp}. To show
 it more clearly, the projections on $x-y$, $x-l$ and $y-l$ planes are
 also plotted in fig \ref{projexp}.

 Fig \ref{l-x-y-exp} illustrates the evolution of a tachyon attached
 to a DGP brane in the phase space $x-y-l$. One clearly see that the orbits with different initial conditions converge to
 one orbit, which is helpful to explain the present amplitude of the cosmological constant. In this converging flow of orbits,
 different initial conditions yield almost the same universe.   But we have proved that it does not exist a strict
 attractor in this system. Thus, it is only a quasi-attractor. Different orbits
 are associated with different initial conditions. Fig \ref{l-x-y-exp} describes the evolution
 of the universe from $s=-1$ to $s=3$. The slow-roll
 parameter $\alpha \approx 8.6\times 10^{-6}<<1$ at the quasi-attractor.
 Therefore, slow-roll is a perfect approximation and the universe is effectively a de Sitter one. The detailed parameters for this
 quasi-attractor are listed as follows: $j=0.01$, $\Omega_{dm0}=0.3$, $\Omega_{r_c}=0.2$.  $\Omega_{dm0}$ and $\Omega_{r_c}$
 are present partitions of the dust matter and geometric term, which
 are defined as
 \be
 \Omega_{dm0}\triangleq\frac{\rho_{dm0}}{\rho_c},
 \en
 and
  \be
 \Omega_{r_c}\triangleq\frac{\rho_{0}}{\rho_c},
 \en
 where $\rho_{dm0}$ labels the present density of dust matter and
 $\rho_c$ denotes the present critical density.

\begin{figure}
\centering
 \includegraphics[totalheight=5.6in, angle=-90]{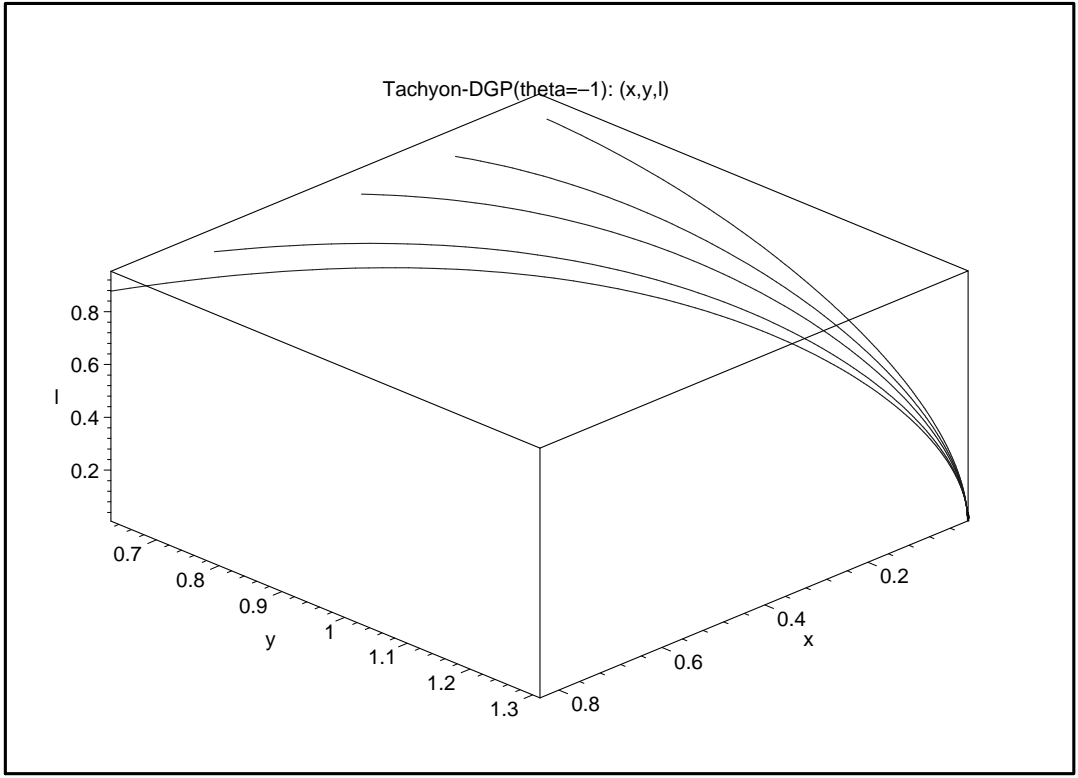}
\caption{The evolution of a tachyon on a DGP in the branch
$\theta=-1$. The different initial conditions for the curves are
$x(s=0)=0.1, x(s=0)=0.08, x(s=0)=0.05, x(s=0)=0.03, x(s=0)=0.01$
from the left to the right, respectively. }
 \label{l-x-y-exp-1}
 \end{figure}

\begin{figure}
\centering
 \includegraphics[totalheight=3in, angle=-90]{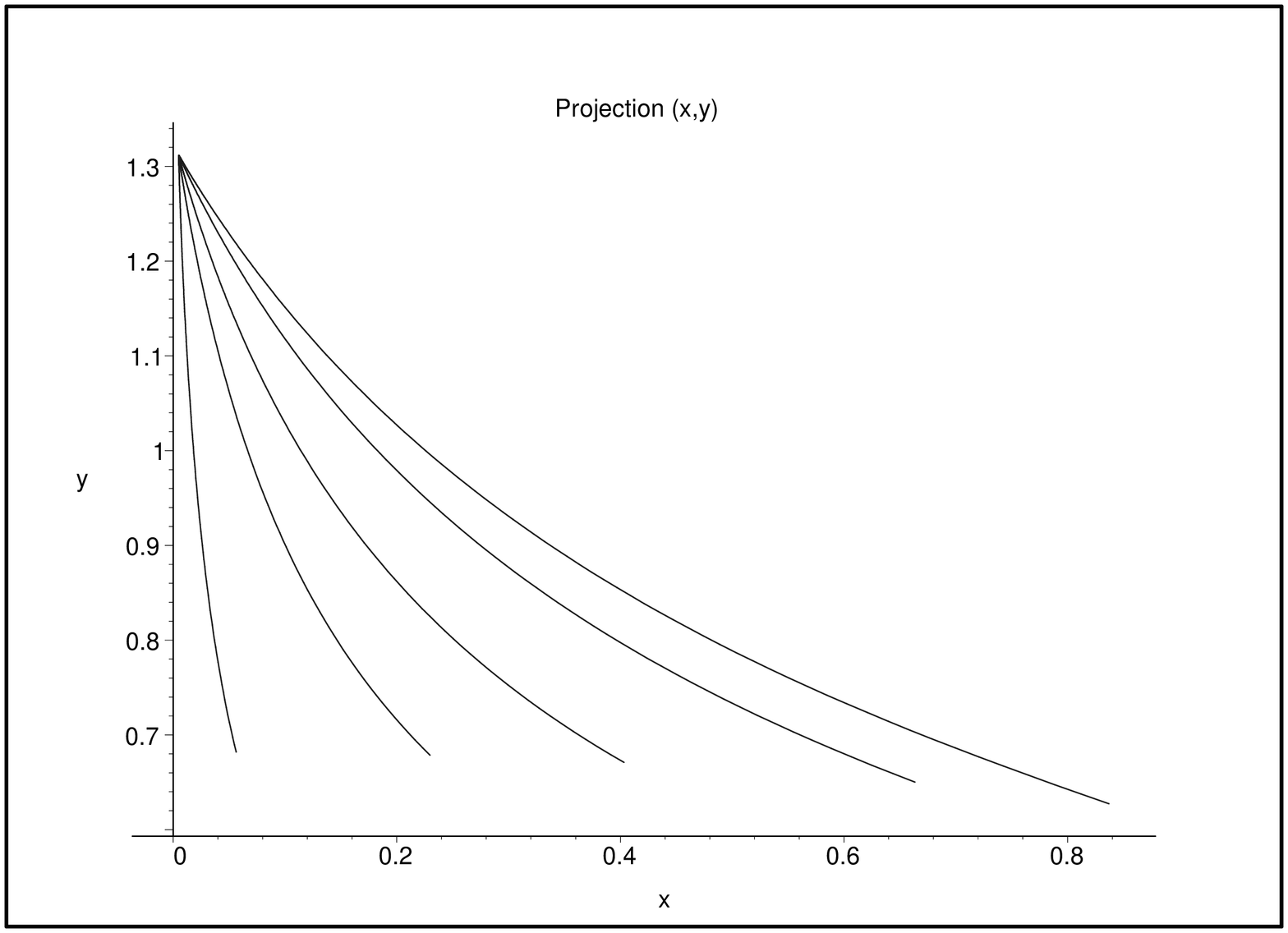}
 \includegraphics[totalheight=3in, angle=-90]{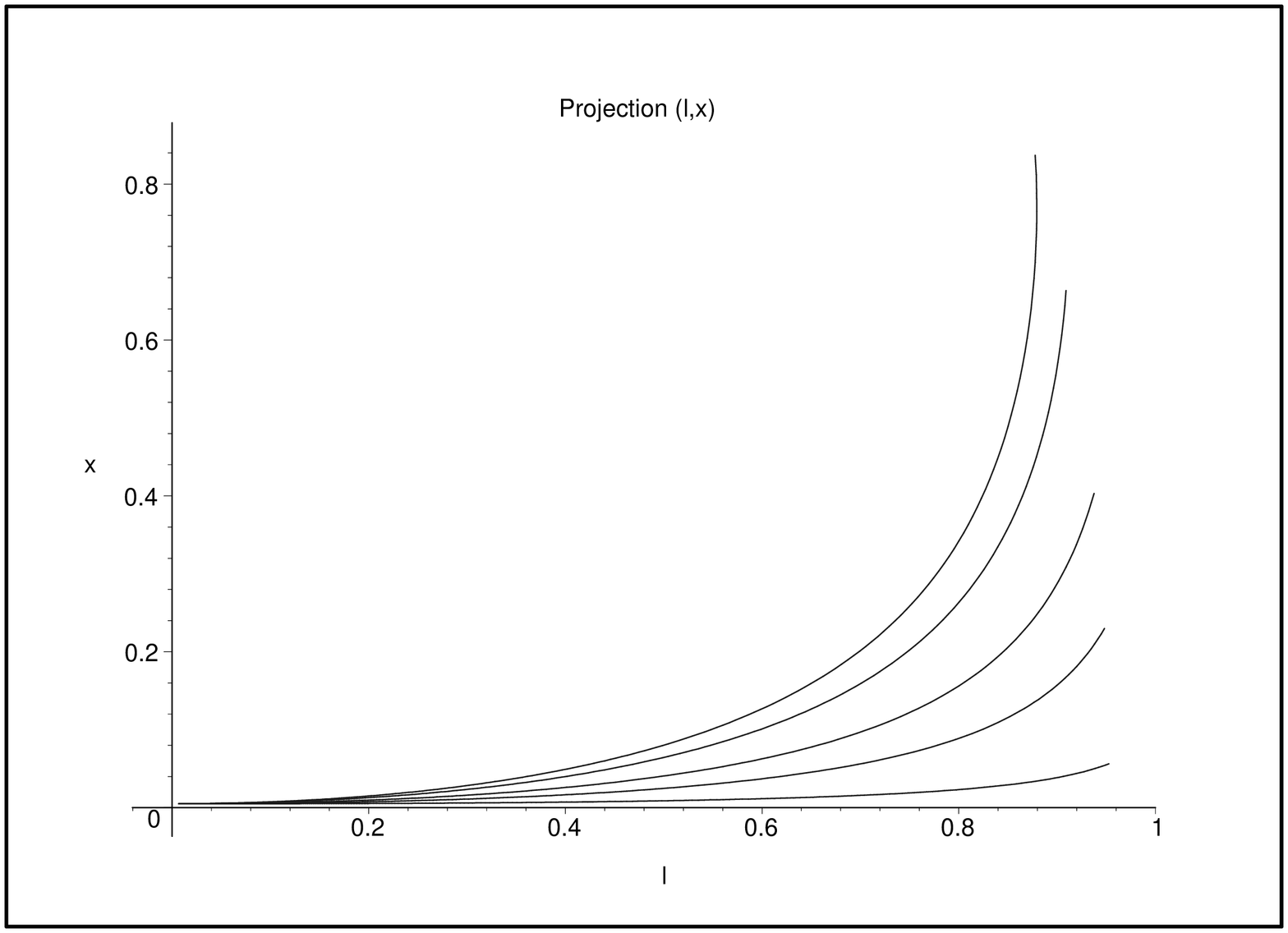}
 \includegraphics[totalheight=3in, angle=-90]{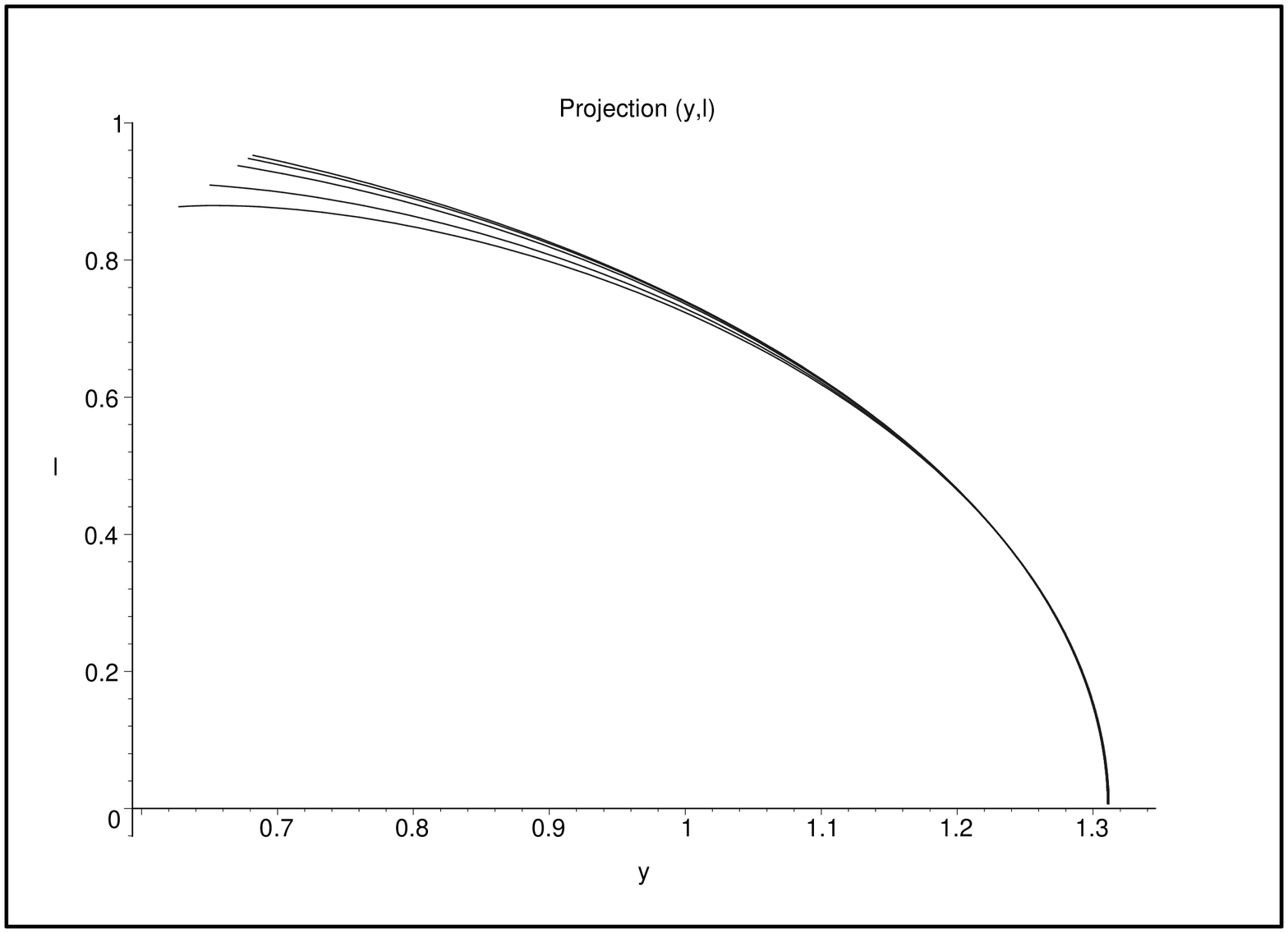}
\caption{The projections of fig \ref{l-x-y-exp-1} on $x-y$, $x-l$,
$l-y$ planes, respectively.}
 \label{projexp-1}
 \end{figure}

 For the branch $\theta=-1$, we have a similar conclusion. We show
 the phase portraits of $x-y-l$ in fig \ref{l-x-y-exp-1} and its
 projections in fig \ref{projexp-1}. Fig \ref{l-x-y-exp-1} describes the evolution
 of the universe from $s=-1$ to $s=3$.  For comparison, the parameters
 in the figs \ref{l-x-y-exp-1} and \ref{projexp-1} are adopted as
 the same of the branch $\theta=1$. From the panel $l-y$ in fig
 \ref{projexp}, one sees that the different curves almost
 coincide, which implies that the phase space is almost reduced
 to a lower dim subspace.

  From fig \ref{l-x-y-exp-1}, one
 sees that the quasi-attractor appears again in the branch
 $\theta=-1$.
 $\alpha \approx 3.2\times 10^{-5}<<1$ at the quasi-attractor, which marks slow-roll of the tachyon. The
 corresponding density and pressure read,

 \be
 \frac{\rho_{T}}{\rho_c}=1.72106,
 \en
 and
 \be
 \frac{p_{T}}{\rho_c}=-1.72101\approx -\frac{\rho_{T}}{\rho_c}.
 \en
 Though it looks the tachyon is a perfect approximation of vacuum,
 we  stress that the density $\rho_{de}$ and $p_{eff}$ in (\ref{contieff}) are different
 from $\rho_{T}$ and $p_{T}$. The evolution of the universe around the quasi-attractor
 is determined by the joint effect of the tachyon and geometric
 contribution, for the dust matter has been completely diluted away.

\begin{figure}
\centering
 \includegraphics[totalheight=1.87in, angle=0]{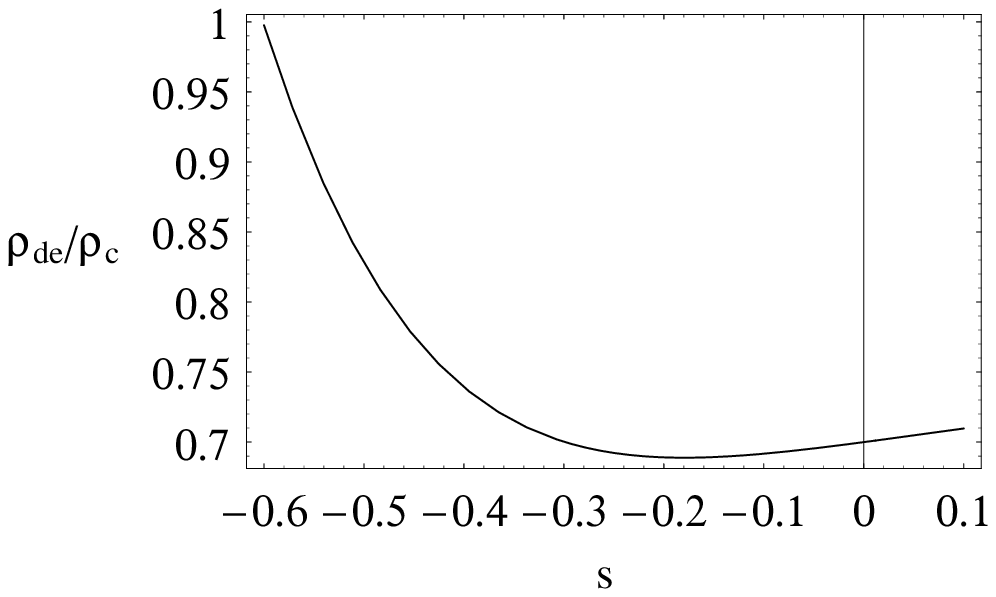}
\includegraphics[totalheight=1.8in, angle=0]{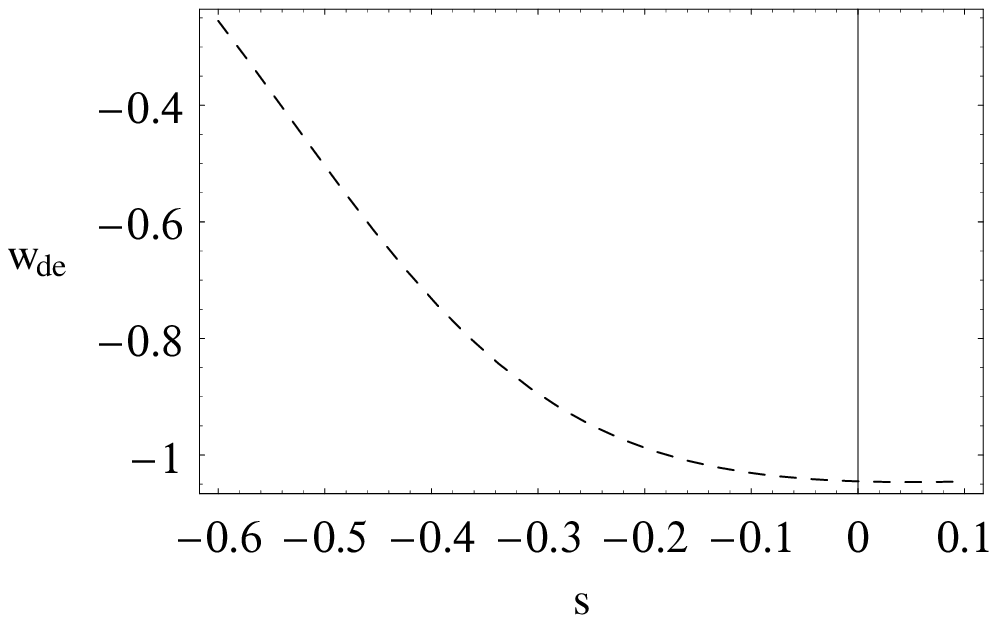}
\caption{Density of virtual dark energy and its EOS in the branch
 $\theta=-1$ for an exponential potential. Left panel: The evolution of the density of virtual dark energy as a
function of $s=\ln a$. Right panel: The evolution of the EOS of the
virtual dark energy as a function of $s$. The present epoch is
denoted by $s=0$.}
 \label{wrhoexp}
 \end{figure}

 So, we further study the behavior of the virtual dark energy, which
 carries the combining effect of the tachyon and geometric term. We
 find that  on the way to the quasi-attractor the crossing $-1$ behavior of
   the EOS of the virtual dark energy will
  appear in the branch $\theta=-1$.  We show a concrete numerical example of this crossing
  behaviors in fig. \ref{wrhoexp}, in which we take the parameter set
  as $j=0.01$, $\Omega_{dm0}=0.3$, $\Omega_{r_c}=0.2$.

 Fig \ref{wrhoexp} explicitly illuminates that the EOS crosses $-1$ at $s\sim
 -0.2$. Also, we plot the evolution of the deceleration parameter $q$. It is one of the most significant parameters from the viewpoint of
  observations, which carries the total
  effects of cosmic fluids. $q$ is defined as,
  \bea
  \nonumber
  q&\triangleq&-\frac{\ddot{a}a}{\dot{a}^2}\\
   &=&-1+\frac{3}{2}\alpha.
   \ena

 \begin{figure}
\centering
\includegraphics[totalheight=1.8in, angle=0]{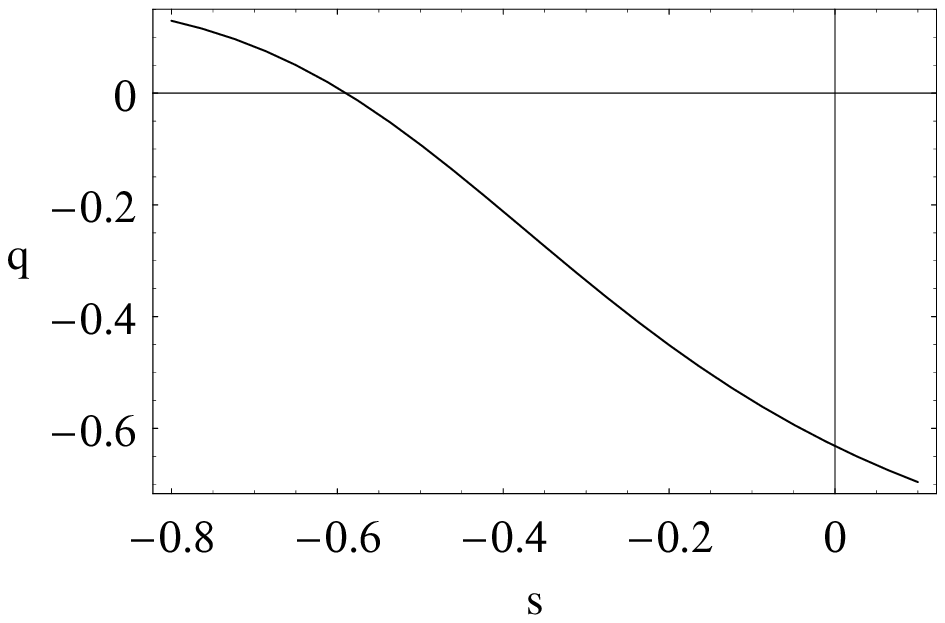}
\caption{The deceleration parameter as a function of $s$
corresponding to fig \ref{wrhoexp}.}
 \label{decexp}
 \end{figure}

 Fig \ref{decexp} illuminates the evolution of the deceleration
 parameter for a tachyon on a DGP in the branch $\theta=-1$ with the
 same parameters in the above figure.

 From Fig \ref{wrhoexp} and \ref{decexp}, clearly, the EOS of effective dark energy
 crosses $-1$. At the same time the deceleration
 parameter is consistent with observations. It is well known that
 the EOS of a single scalar in standard GR never
 crosses the phantom divide. Therefore, the induced term, through the ``energy density" of $r_c$,
 $\rho_0$, plays a critical role in this crossing.

 For the branch $\theta=1$, this crossing behavior is impossible. We
 demonstrate this point by using (\ref{wde}) and the explanation following it.
  In the branch $\theta=1$, the virtual dark energy density
  $\rho_{de}$ in (\ref{rhode}) becomes,
 \be
  \rho_{de}=\frac{V(T)}{\sqrt{1-\dot{T}^2}}+\rho_0+\rho_0
 (1+\frac{2(\frac{V(T)}{\sqrt{1-\dot{T}^2}}+\rho_{dm})}{\rho_0})^{1/2}.
  \en
  Clearly, every term in RHS of the above equation is decreasing in
  an expanding universe ($\rho_T$ will decrease since its $w>-1$,
  and $\rho_{dm}$ decreases with the scale factor). This conclusion is
  unchanged for the $\theta=1$ branch of DGP with an essence whose $w>-1$ and dust matter
  confined to it. It is also independent of the concrete form of the
  potential (for an positive potential).
\subsection{Inverse power law potential}
 The discussions of this subsection is
 parallel to  the last subsection.

 Exponential potential is an important case in the standard model. Some researches imply
 that  an analogy of exponential potential in
 the context of tachyon dynamics is inverse square potential.
 Here we set
 \be
 V=\frac{4A^2}{3T^2},
 \en
 where $A$ is a constant.
  The dynamics of the universe can be described by the following
  dynamical system with the dimensionless variables $x, y, l, b$,
 \bea
 \label{1}
 x'&=&{3}(1-x^2)(-x+\frac{\mu}{A}y),\\
 \label{2}
  y'&=&\frac{3}{2}\alpha y-\frac{3\mu}{2A} xy^2,\\
 \label{3}
   l'&=&\frac{3}{2}\alpha l-\frac{3}{2}l,\\
 \label{4}
  b'&=&\frac{3}{2}\alpha b,
 \ena
 where the definition of $\alpha$ is the same as in (\ref{alpha}).
 Similarly, we can prove that there does not exist a strict critical
 point in this system. Note that a critical point in this system must be a de Sitter one if it exists because of the
 definition of the variable $b$. If we define different
 variables, the result may become different.

  We see that neither for the case of exponential potential nor the
  inverse power law potential the attractor does not exist in the
  tachyon-DGP system. In fact, it was shown that for a positive potential the Born-Infled
  type scalar has a critical point only if it has a nonvanishing
  minimum, which corresponds to the de Sitter attractor attractor of
  the system \cite{haoli}. Our results can be treated as examples of the above
  conclusion, for no minimum appears at an exponential potential or an inverse power law potential.

 Like the case of an exponential potential, a quasi-attractor appears again in this
 system, which suggests the universality of the quasi-attractor
 behavior.

 Following the discussions of the last subsection, we  plot 3-dim phase portraits in the subspace $x-y-l$ in the branch $\theta=1$ and $\theta=-1$ respectively,
 and their the projections on $x-y$, $x-l$ and $y-l$ planes.

 \begin{figure}
\centering
 \includegraphics[totalheight=5.6in, angle=-90]{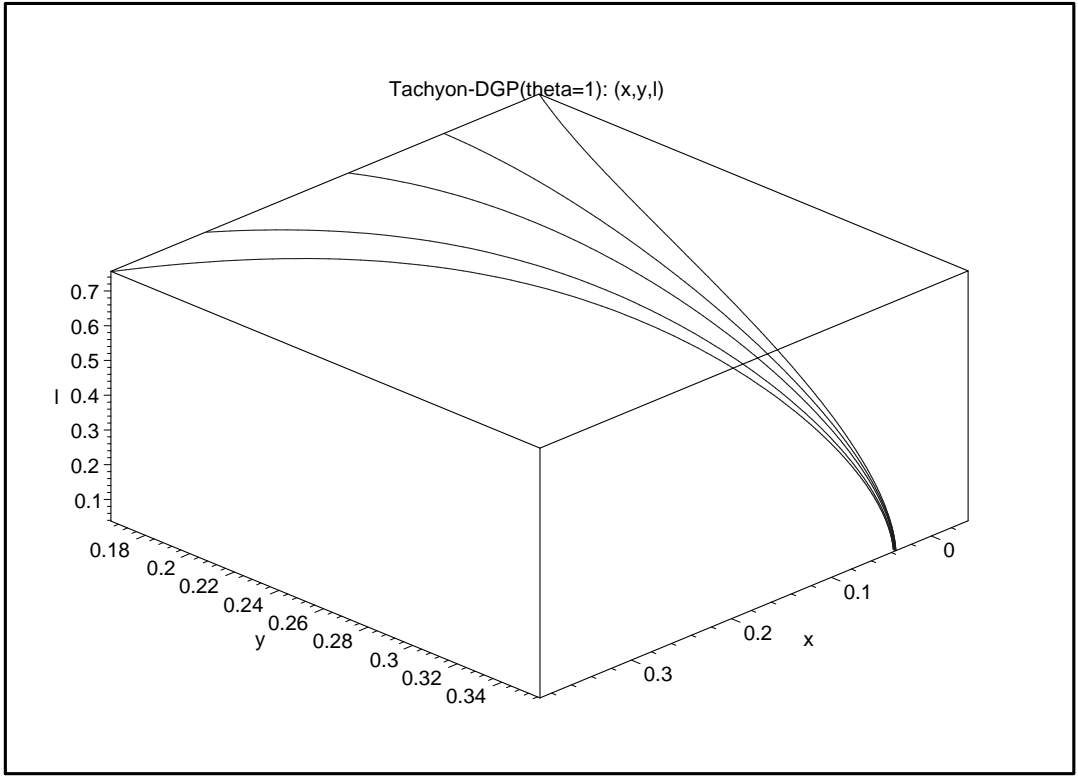}
\caption{The evolution of a tachyon on a DGP in the branch
$\theta=1$.  The parameters used for this
 figure are  $\mu/A=0.1$, $\Omega_{dm0}=0.3$, $\Omega_{r_c}=0.2$.   The different initial conditions for the curves are
$x(s=0)=0.1, x(s=0)=0.08, x(s=0)=0.05, x(s=0)=0.03, x(s=0)=0.01$
from the left to the right, respectively.}
 \label{l-x-y-pow}
 \end{figure}

  \begin{figure}
\centering
 \includegraphics[totalheight=3in, angle=-90]{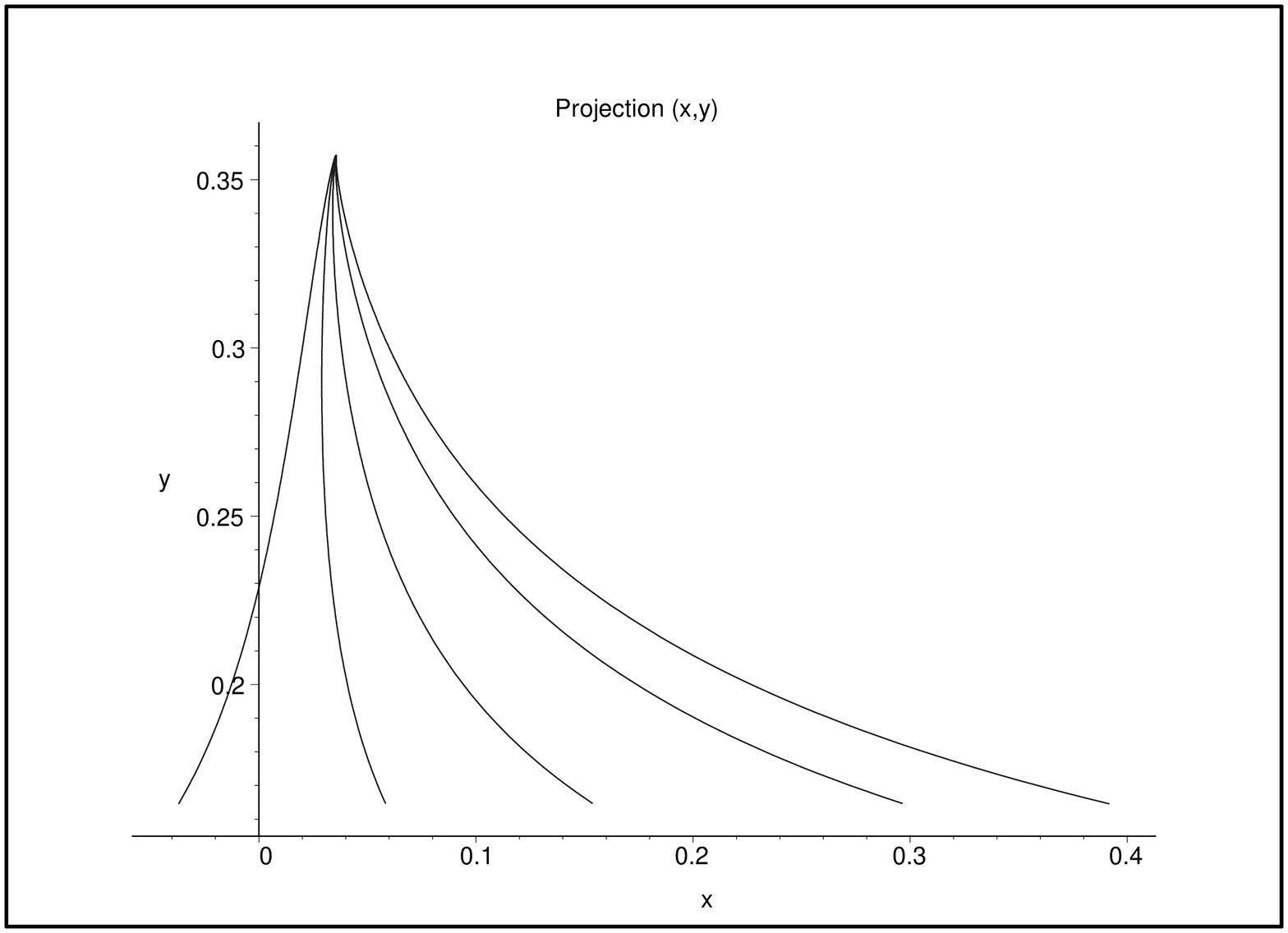}
 \includegraphics[totalheight=3in, angle=-90]{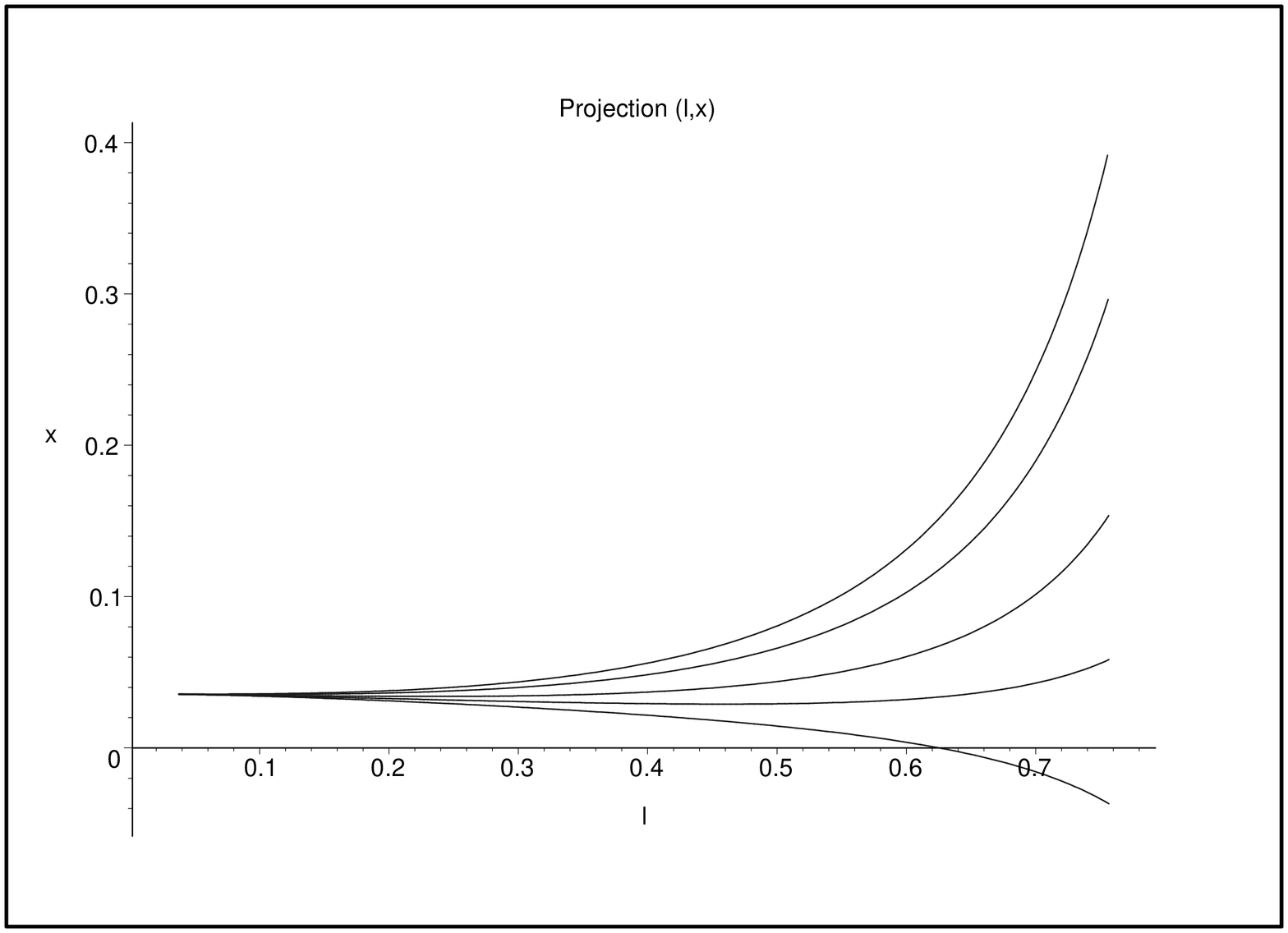}
 \includegraphics[totalheight=3in, angle=-90]{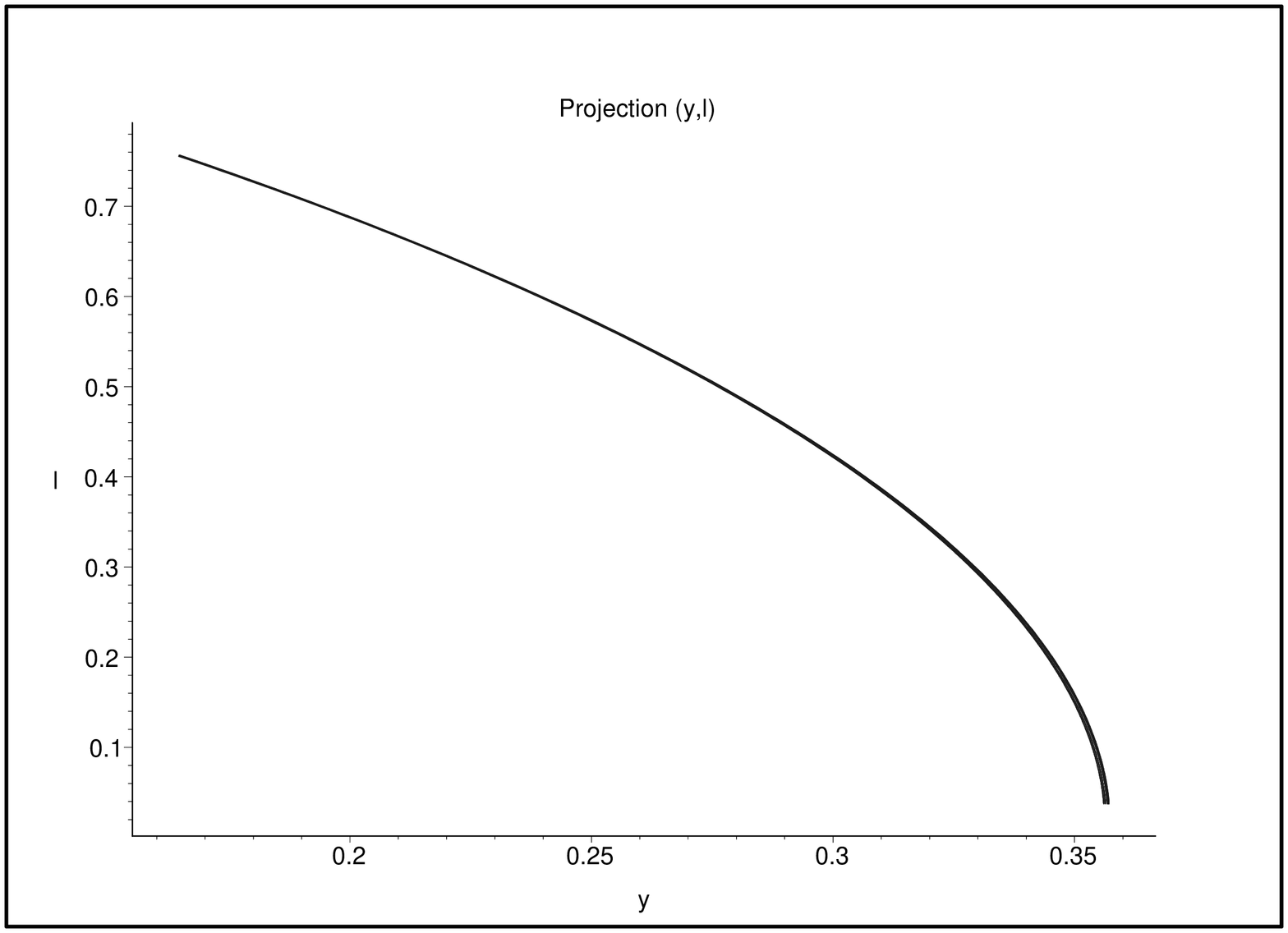}
\caption{The projections of fig \ref{l-x-y-pow} on $x-y$, $x-l$,
$l-y$ planes, respectively.}
 \label{projpow}
 \end{figure}

 Fig \ref{l-x-y-pow} displays the evolution of a tachyon attached
 to a DGP brane in the phase space $x-y-l$. The converging orbits indicate the same
 final state for different initial conditions. However, we showed that there is no strict
 attractor in this system. It is just a quasi-attractor.  Different orbits
 correspond to different initial conditions.  In this case, the slow-roll
 parameter $\alpha \approx 5.5\times 10^{-4}<<1$ at the quasi-attractor.
 Therefore, the case is very similar to what happens in slow-roll inflation and the universe is
 effectively a de Sitter one. Fig \ref{l-x-y-pow} and the following figure \ref{l-x-y-pow-1} describe the evolution
 of the universe from $s=-0.6$ to $s=3$.

 For the branch $\theta=-1$, we have almost the same conclusion.
 The phase portraits of $x-y-l$ and its projections are displayed in fig \ref{l-x-y-pow-1} and \ref{projpow-1}, respectively.
  We set the same parameters
 in the figs \ref{l-x-y-pow-1} and \ref{projpow-1} as
 in the branch $\theta=1$.

   \begin{figure}
\centering
 \includegraphics[totalheight=5.6in, angle=-90]{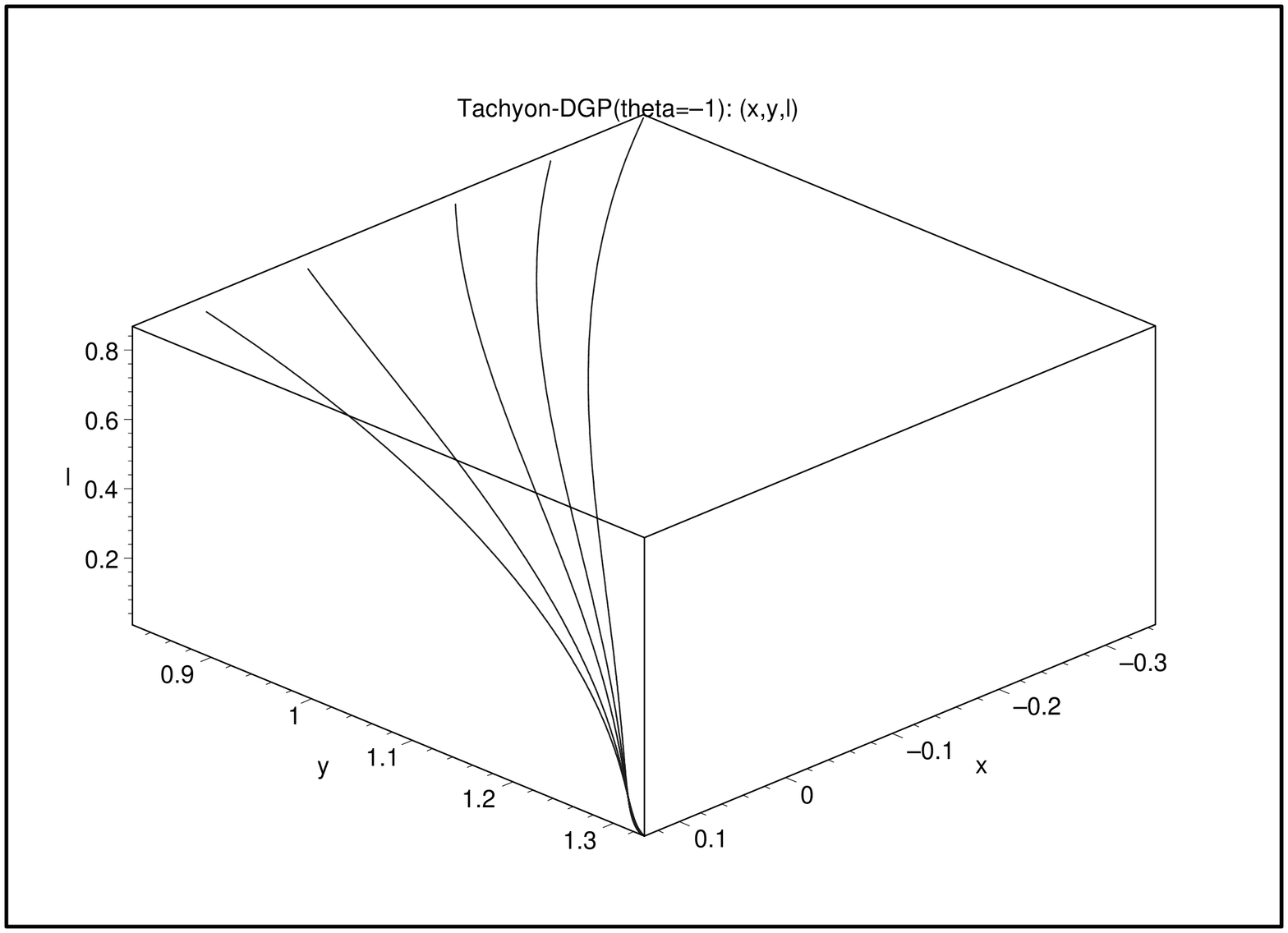}
\caption{The evolution of a tachyon on a DGP in the branch
$\theta=-1$, where the parameters and initial conditions are the
same as in the branch $\theta=1$.}
 \label{l-x-y-pow-1}
 \end{figure}

  \begin{figure}
\centering
 \includegraphics[totalheight=3in, angle=-90]{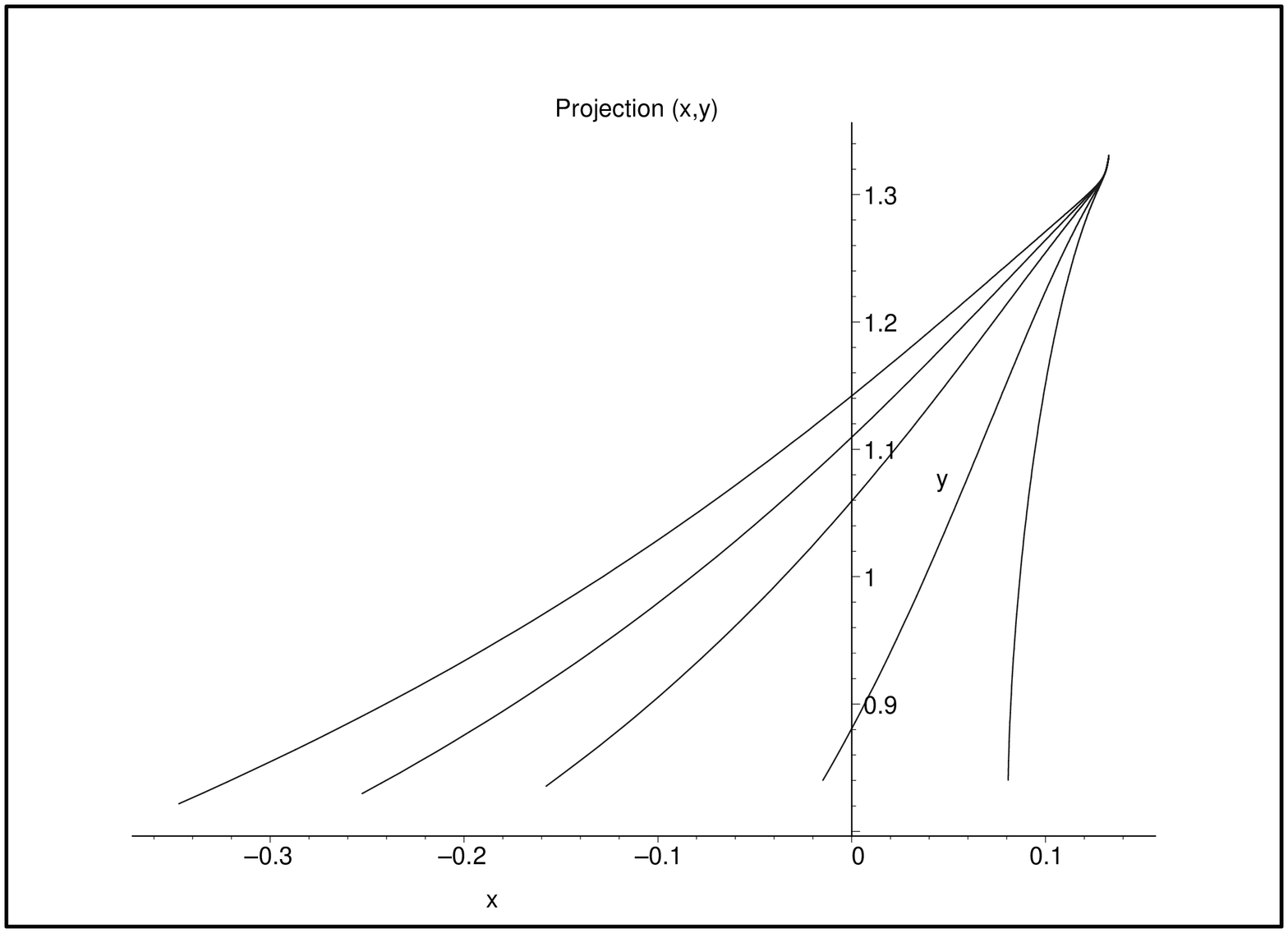}
 \includegraphics[totalheight=3in, angle=-90]{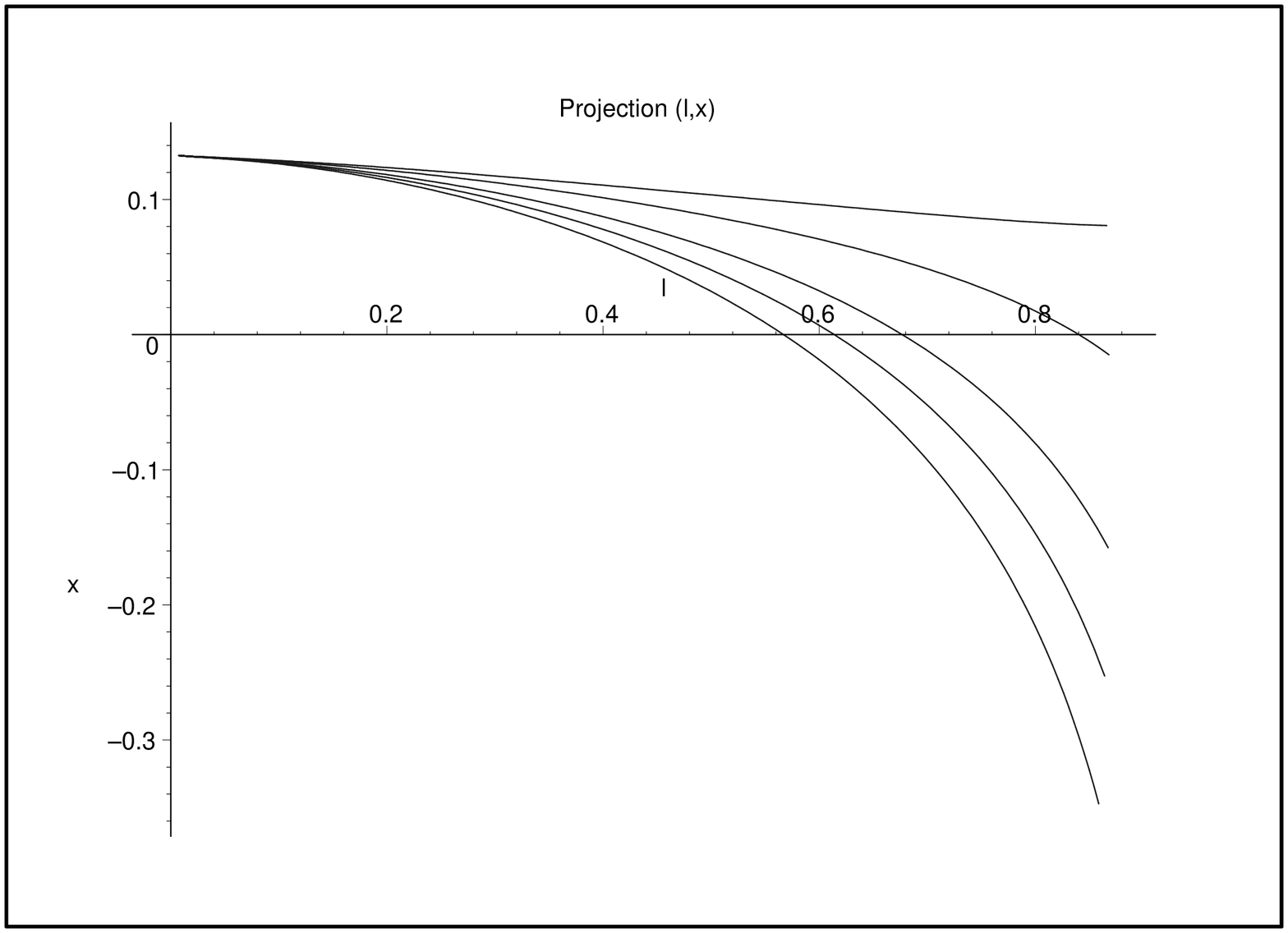}
 \includegraphics[totalheight=3in, angle=-90]{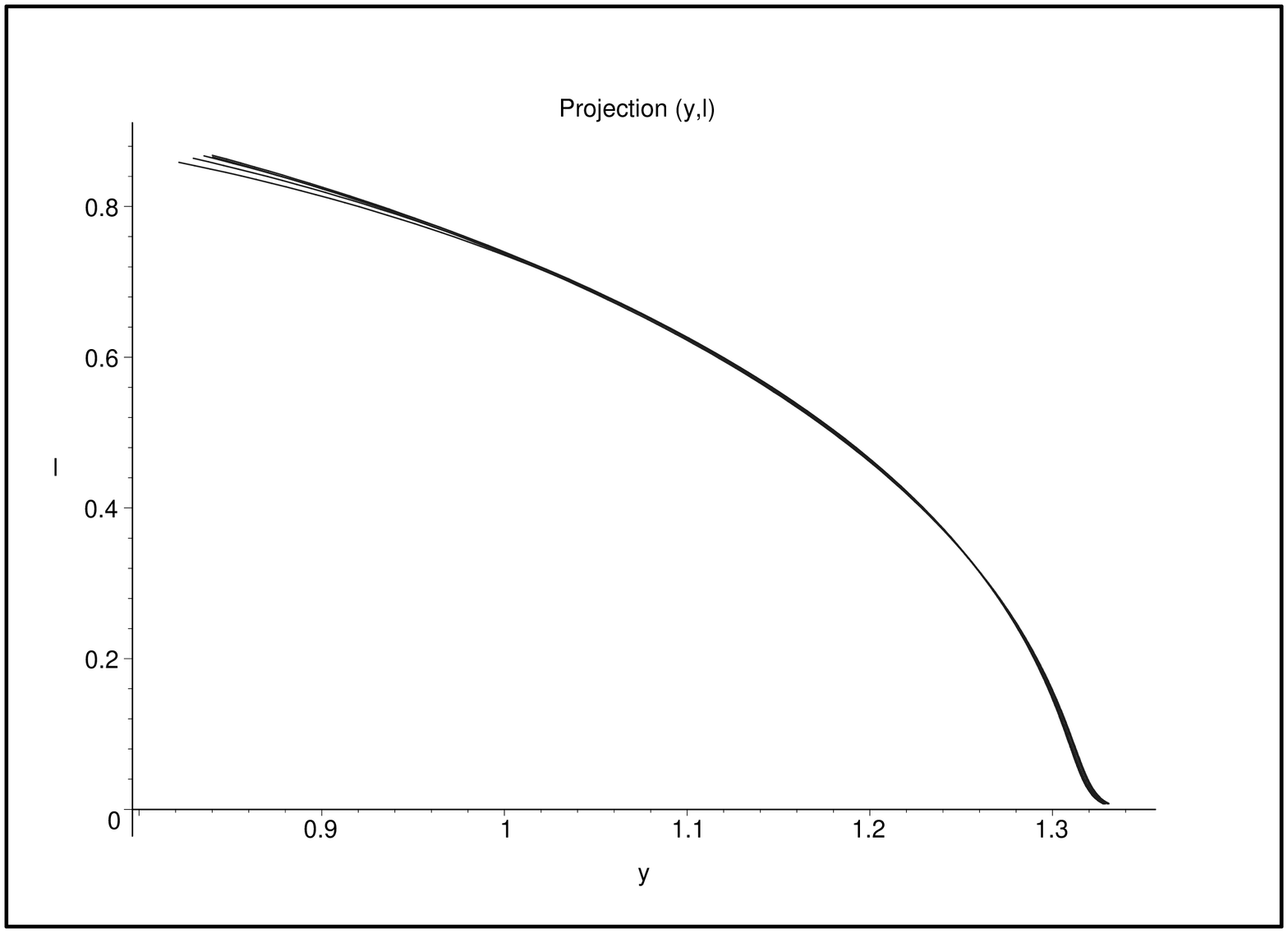}
\caption{The projections of fig \ref{l-x-y-pow-1} on $x-y$, $x-l$,
$l-y$ planes, respectively.}
 \label{projpow-1}
 \end{figure}

 From fig \ref{l-x-y-pow-1}, the quasi-attractor appears again as we expected. In branch $\theta=-1$,
 $\alpha \approx 0.029<<1$ at the quasi-attractor. The
 corresponding density and pressure read,

 \be
 \frac{\rho_{T}}{\rho_c}=2.20,
 \en
 and
 \be
 \frac{p_{T}}{\rho_c}=-2.15\approx -\frac{\rho_{T}}{\rho_c}.
 \en
 We see that tachyon  finally evolves as cosmological constant. However, the evolution of the universe is not
  determined by the tachyon only, but by the joint effect of the tachyon and geometric
  term. In the following we will study the evolution of the virtual
  dark energy, which carries the total effect of the tachyon and the
  geometric effect and determines the destiny of the universe.

\begin{figure}
\centering
 \includegraphics[totalheight=1.8in, angle=0]{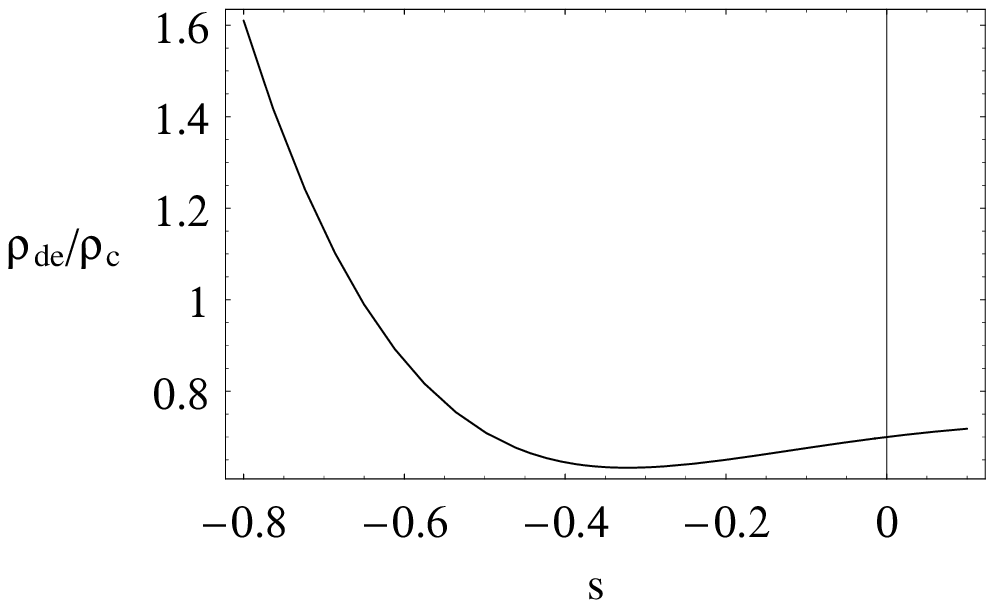}
\includegraphics[totalheight=1.8in, angle=0]{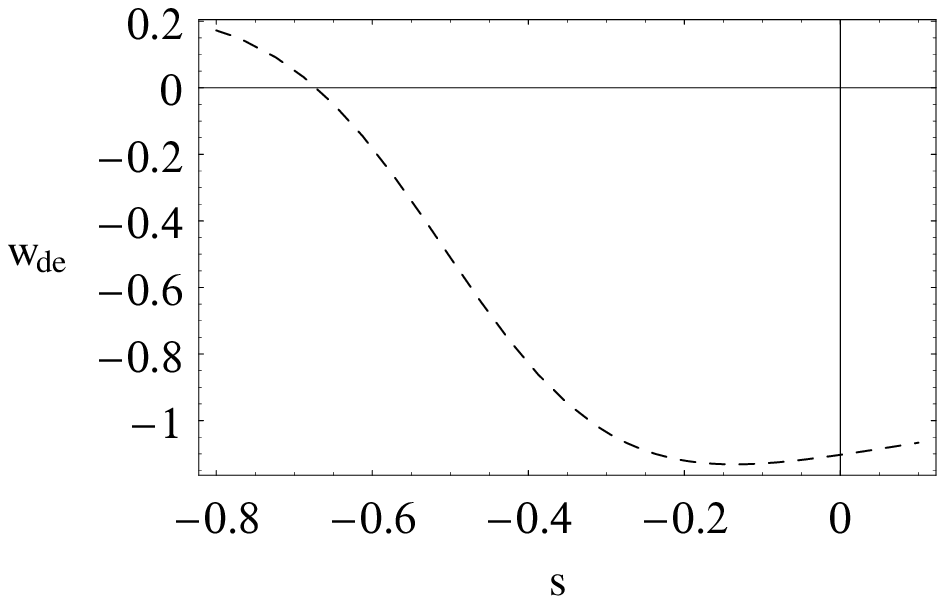}
\caption{Density of virtual dark energy and its EOS in the branch
 $\theta=-1$ for an inverse power law potential. Left panel: The evolution of the density of virtual dark
energy as a function of $s$. Right panel: The evolution of the EOS
for the virtual dark energy as a function of $s$.}
 \label{wrhopow}
 \end{figure}

   We
 find that  when the universe is approaching the quasi-attractor the EOS of the virtual dark energy
 can cross the phantom divide in the branch $\theta=-1$.  A concrete numerical example of this crossing
  behaviours is displayed in fig. \ref{wrhopow}, in which we take the parameter set
  as $\mu/A=0.25$, $\Omega_{dm0}=0.3$, $\Omega_{r_c}=0.2$.

 Fig \ref{wrhopow} clearly displays that the EOS crosses $-1$ at  $s\sim
 -0.3$. At the same time we plot the corresponding  deceleration parameter in fig \ref{wrhopow},
 which is determined by the total fluids in the universe.

\begin{figure}
\centering
\includegraphics[totalheight=1.8in, angle=0]{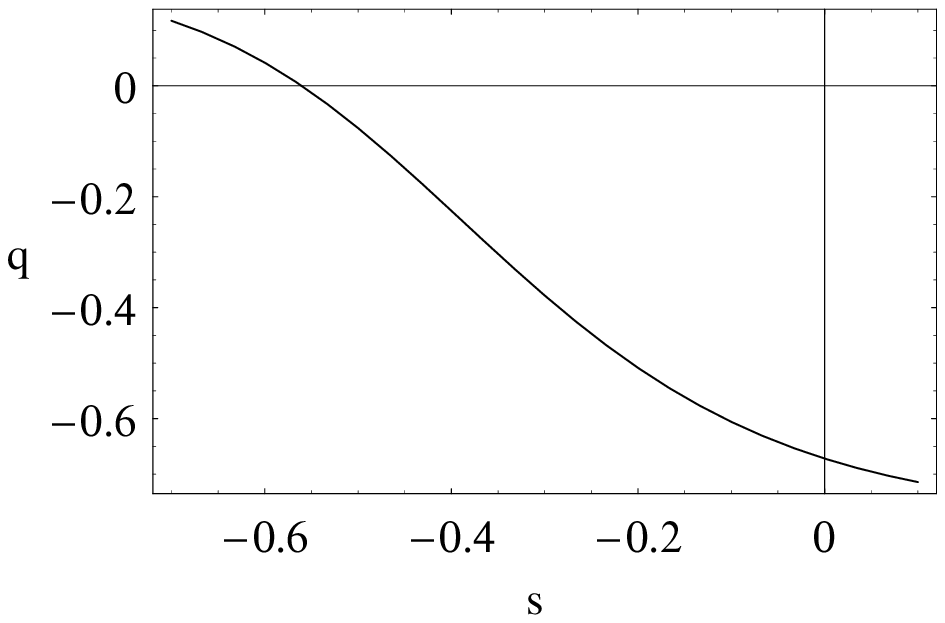}
\caption{The deceleration parameter as a function of $s$
corresponding to fig \ref{wrhopow}.}
 \label{decepow}
 \end{figure}

 Fig \ref{decepow} illuminates the evolution of the deceleration
 parameter for a tachyon on a DGP in the branch $\theta=-1$ with the
 same parameters in the above figure.

 From fig \ref{wrhopow} and \ref{decepow}, one sees  the EOS of effective dark energy
 crosses $-1$. Also, the deceleration
 parameter is consistent with observations.

 From the conclusion which we obtained in the last subsection, the crossing does not appear in the $\theta=1$ branch.

\section{Conclusions and discussions}

 In this article, the dynamics of a tachyon attached to a DGP brane
 is studied.  Two kinds of potentials of the tachyon field,
 exponential potential and inverse power law potential, are
 explored, respectively.

 In the investigation of tachyon-DGP, we find the quasi-attractor
 behavior. Traditionally, if a dynamical system does not permit
 critical points, we just stop,  imagining that the orbits in this
 system must be disordered and never converge. However, we find that
 the orbits with different initial conditions  converge even though
 there is no real critical points. This quasi-attractor is full of vitality,
  which can appear in both of the two branches, and for both of the
 two kinds of  potentials. The converging evolution of the orbits in the
 phase portraits offers a new view on the cosmological constant
 problem and coincidence problem. The analytical work for this quasi-attract behavior need to do in the future.

 In the branch $\theta=-1$, we find that the EOS of the virtual dark energy,
 which is caused by the tachyon and geometric term, can cross the
 phantom divide for both of exponential potential and inverse power
 law potential. This provides a new theoretical possibility for the
 extraordinary observation of dark energy. We find that the
 geometric term plays a significant role in this crossing.
 Contrarily, the crossing behavior do not appear in the branch
 $\theta=1$.


{\bf Acknowledgments}: H Noh was supported by Grant Mid-Career Research program through NRF funded by MEST
(2010-0000302). We thank the anonymous reviewer for several valuable suggestions.

\end{document}